\documentclass[10pt,letterpaper]{article}
\usepackage[top=0.85in,left=2.75in,footskip=0.75in,marginparwidth=2in]{geometry}
\usepackage[utf8]{inputenc}
\usepackage{cite}
\usepackage{nameref,hyperref}
\usepackage[right]{lineno}
\usepackage{microtype}
\DisableLigatures[f]{encoding = *, family = * }
\raggedright
\usepackage{changepage}
\usepackage[aboveskip=1pt,labelfont=bf,labelsep=period,singlelinecheck=off]{caption}
\makeatletter
\renewcommand{\@biblabel}[1]{\quad#1.}
\makeatother
\usepackage{lastpage,fancyhdr,graphicx}
\usepackage{epstopdf}
\fancyheadoffset[L]{2.25in}
\fancyfootoffset[L]{2.25in}
\usepackage{color}
\usepackage{multirow}
\usepackage{nameref}
\usepackage{amssymb, amsfonts, amsmath}

\usepackage{array}
\usepackage{stfloats}
\usepackage{booktabs, tabularx}
\usepackage{subfig}
\usepackage{alphalph}
\usepackage{ragged2e,placeins}
\graphicspath{{../}}

\usepackage{rotating}

\usepackage[table, dvipsnames]{xcolor}
\definecolor{light-gray}{HTML}{E5E4E2}

\usepackage{setspace}
\usepackage{nicefrac}

\usepackage{lineno,hyperref}
\modulolinenumbers[5]

\definecolor{Gray}{gray}{.25}

\usepackage{sidecap}

\usepackage{wrapfig}
\usepackage[pscoord]{eso-pic}
\usepackage[fulladjust]{marginnote}
\reversemarginpar

\begin{document}
\vspace*{0.35in}

% title goes here:
\begin{flushleft}
{\Large
\textbf\newline{Evaluation of Biases in Self-reported Demographic and Psychometric Information: Traditional versus Facebook-based Surveys}
}
\newline
% authors go here:
\\

Kyriaki Kalimeri\textsuperscript{1,$\ast$}
%\fntext[myfootnote]{Corresponding Author. E-mail: kalimeri$@$ieee.org}
Mariano G. Beir\'{o}\textsuperscript{2}
Andrea Bonanomi\textsuperscript{3} 
Alessandro Rosina\textsuperscript{3}
Ciro Cattuto\textsuperscript{1}
\\

\bf{1} Data Science Laboratory,  ISI Foundation,  Turin,  Italy
\\
\bf{2} Universidad de Buenos Aires, Facultad de Ingenier\'{i}a, INTECIN (CONICET), CABA, Argentina
\\
\bf{3} Universit\`{a} Cattolica del Sacro Cuore (UNICATT), Milan, Italy
\\

$\ast$  Corresponding author. E-mail: kalimeri$@$ieee.org
\end{flushleft}

\justify

\begin{abstract}
Social media in scientific research offer a unique digital observatory of human behaviours and hence great opportunities to conduct research at large scale answering complex sociodemographic questions.
%Despite the great potentials, there is an ever-growing concern about inherent biases hard to identify and quantify.
We focus on the identification and assessment of biases in social media administered surveys.
This study aims to shed light on \textit{population}, \textit{self-selection} and \textit{behavioural} biases, empirically comparing the consistency between self-reported information collected traditionally versus social media administered questionnaires,  including demographic and psychometric attributes.
We engaged a demographically representative cohort of young adults in Italy (approximately 4,000 participants) in taking a traditionally administered online survey and then, after one year, we invited them to use our ad hoc Facebook application (988 accepted) where they filled in part of the initial survey.
%hypothesis testing
%With this design we assessed two major aspects; firstly,  \textit{population} and \textit{self-selection} biases in the demographic or psychometric attributes, during the recruitment phase of our study, or the platform itself; and secondly, the \textit{behavioural} biases due to the different context in which the questionnaire is administered.
We assess the statistically significant differences indicating \textit{population}, \textit{self-selection}, and \textit{behavioural} biases due to the different context in which the questionnaire is administered.
Our findings suggest that surveys administered on Facebook do not exhibit major biases with respect to traditionally administered surveys neither in terms of demographics, nor personality traits.
Loyalty, authority, and social binding values were higher in the Facebook platform, probably due to the platform's intrinsic social character.
We conclude, that Facebook apps are valid research tools for administering demographic and psychometric surveys provided that the entailed biases are taken into consideration.
We contribute to the characterisation of Facebook apps as a valid scientific tool to administer demographic and psychometric surveys, and to the assessment of population, self-selection, and behavioural biases in the collected data.
%we contribute to the characterization of Facebook apps as a scientific tool to administer surveys and collect demographic and psychometric data
%We point out the statistically significant differences that emerged in demographic and psychometric attributes that should be taken into consideration by the researchers when using Facebook as a research tool.\\

%\keywords{Demography, Psychometrics, Facebook, survey, self-selection bias, recruitment bias, social media, self-reporting bias, Personality, Moral Foundations}
\end{abstract}

\section{Introduction}
%Like never before in the human history, we have access to a unique digital observatory of human behaviours depicted, among others sources, on social media data.
Scientific research, and in particular cognitive and social sciences, are going through a revolution in light of the immense possibilities that arise from the availability to observe real-world human behaviours at a large scale depicted, among others sources, on social media data.
Computational social science is exactly the research area where novel computational methods are used to answer questions about society \cite{Lazer2009}. 
Demography, as the science of human populations, is lying at the intersection point of the social, behavioural, and statistical sciences,  encompassing a variety of issues,  among which, population and development, environmental, equity, transnationalism, migration, as well as their implications on life events. 
%Digital datasets offer rich, fine-grained information at individual-level, enriching the traditional data sources with the possibility to address complex challenges like illegal immigration at a population scale.
% and is one of the disciplines where social media data can have a direct impact.
The demographic community is paying an ever-growing attention to social media platforms as they offer an alternative, complementary view of the society providing along with rich, fine-grained information at individual-level at a population level.
Integrated with data obtained via traditional sources such as census or surveys, social media data provided with the opportunity to address research questions timely and in greater scale.
%, shedding light also on complex sociological and psychological dimensions tied to certain demographic phenomena.
%This new perspective is gaining popularity is the opportunity to integrate demographic data from social surveys, such for example individual characteristics and life events, with social media information regarding psychological and behavioural attributes.  

This new perspective is gaining popularity since it offers the possibility of shedding light on complex psychological dimensions, which are often fundamental explanatory factors related to many sociological phenomena directly tied to demographic issues. 
The importance of understanding these factors is already present in the second demographic transition (SDT) \cite{Lesthaeghe2007}, which entails a macro-level view of societal development, where the values, behaviours, and psychological attributes play an important role and may vary across contexts.
The position taken closely relates demographic issues to Abraham Maslow's theory \cite{Maslow1954}, according to which
well-being occurs to the extent people can freely express their inherent potentials, while values, motivation and personality are given a central role.
Over the last years, an increasing body of studies is dealing with the relationship between psychological well-being and demographic issues such for example: (i) leaving the parents' home \cite{Parr2010}, (ii) family formation and dissolution \cite{Moor2012}, (iii) childlessness \cite{Bernardi2014}, (iv) living alone \cite{Ho2015}, (v) the elderly condition \cite{Teerawichitchainan2015}, and (vi) mobility \cite{Morrison2016}. 
Social media data can act as proxy to these attributes; not only they can be employed for the estimation of population in regions where official records are inaccessible, but also for the assessment of psychological attributes \cite{Schwartz2013,Golbeck2011,Kosinski2013}, and social phenomena of great importance such as migration patterns \cite{Zagheni2014}, global mobility patterns \cite{Hawelka2014} and misinformation \cite{Bessi2014},
 health monitoring \cite{Ginsberg2009} and epidemic spreading \cite{Vespignani2009}.
The potentials of social media are demonstrated in a series of applications such as for example crisis response \cite{Imran2015} and
deployment of resources during health emergencies \cite{Vespignani2009}.
Political mobilisation, too, is influenced by the way people use social media; Margetts et al.~\cite{Margetts2015} drawing on large-scale data generated from the Internet showed how mobilisations that succeed are unpredictable, unstable, and often unsustainable.
Social media influence citizens deciding whether or not to participate depending on their personality types \cite{Margetts2015}, shaping the networked public sphere and facilitating communication between communities with different political orientations \cite{Conover2011}.

In the demographic research field, the use of digital sources brought several new challenges and potential benefits which, however, require the development of new techniques that consider the potential biases and representativeness of the data \cite{Cesare16,Zagheni2015}.
For instance, Twitter emerged as the predominant platform to study social phenomena, such as predicting elections results \cite{Bermingham2011} and political behaviour in general \cite{Digrazia2013},  due to key features, for instance, direct, timely, and brief communication as well as the easiness in obtaining data.
Recently, researchers are exploring the possibilities offered by Facebook (FB) to address complex societal and demographic questions for instance gender gap tracking \cite{Fatehkia2018}, monitoring \cite{Zagheni2017} and assimilation of migrants \cite{Dubois2018}, unemployment \cite{Burke2013,Liorente2015,Bonanomi2017}, and measuring labor markets \cite{Antenucci2014}.

Despite the plentifulness and great potentials of the social media data, all data sources come with their own biases and limitations \cite{Reips2002,Pew2018,Araujo2017}; defining and quantifying them is still a major challenge \cite{Baeza2018}.
Internet obtained data provide, undoubtedly, a qualitative shift in the scale, scope and depth of possible analysis, issues about the quality and biases arose quickly~\cite{gosling2004should,Tufekci2014,Olteanu2016}.
The Twitter platform for example entails pronounced sample \cite{Tufekci2014,Metaxas2011} and content biases \cite{Wu2011,Baeza2018}. 
Olteanu et al. \cite{Olteanu2016}, provided an in-depth survey on the methodological limitations and pitfalls, as well as ethical boundaries and unexpected consequences which are often overlooked. 
Apart from well-studied data quality issues, such as sparsity \cite{Baeza2013}, representativity \cite{Ruths2014}, and noise \cite{Salganik2017}, social media data entail biases that are much more difficult to quantify. 
In line with the scheme proposed by Olteanu et al. \cite{Olteanu2016},  we place the focal point exclusively on the following three types of biases;
(i) \textit{population}, differences in demographics or other user characteristics between a population of users represented in a dataset or platform and a target population,
(ii) \textit{self-selection}, which may occur when relying on self-reports on a certain aspect which may be biased due to what users chose to report, when they chose to do it, and how they chose to do it, and
(iii) \textit{behavioural},  due to differences in user behaviour across platforms or contexts.
%Def.: Behavioral Biases. Differences in user behavior across platforms or contexts, or across users represented in different datasets.
% For instance it is observed that users of different traits or demographics use social platforms mechanisms differently \cite{Olteanu2016,Hong2011}, hence, social media data as proxies of personal traits or demographics may vary in reliability.
%
Biases can occur at every stage of the pipeline \cite{Olteanu2016}; here, we frame our study employing the total survey error theory (TSE) \cite{Weisberg2009} which forces attention to both variance and bias terms.
The TSE theory classifies the biases, under examination in this study, to the system of errors occurring throughout the survey process \cite{Groves2010,Saris2014} and can be summarised in (i) sampling error, due to respondents that do not have a Facebook account,
(ii) non-respondent error, due to participants that chose not to participate in the Facebook questionnaire, and (iii) measurement error, due to within-individual variability in responding across different contexts, respectively.

%One of the most common respondent related error sources in self-report studies is \textit{self-reporting bias}, given by the fact that the respondents' subjectivity might skew their answers.
%Regarding the objectives of this study, the errors tied to the biases sources under examination can be classified into:
%(i) Sampling error, Inherent to the fact of addressing a sample from the population of interest;
%(ii) Coverage error, the fact that the addressed population sample does not represent the population of interest;
%(iii) Self-selection error, the fact that some of the addressed respondents decide not take the survey.

To the present moment, there is only a limited body of work comparing online and social media data to traditional media sources as pointed out by a recent study \cite{Schober2016} which are exactly the major contribution of this study. 
We focus on assessing the consistency of results obtained when comparing traditional and Facebook administered surveys with respect not only to the demographic attributes but also the psychometric ones.
We opted for the Facebook platform based on a few key criteria; first and foremost, its popularity \cite{Pew2018}, which permits the communication with a wide audience previously inaccessible; and secondly, the possibility of comparing surveys with observational digital behavioural data, which is of extreme interest to the computational social science field.

Our experimental design consists of two phases;  initially, a survey was administered in a traditional online manner and subsequently,  the same cohort was invited to use an ad hoc Facebook application\footnote{\href{http://likeyouth.org}{Likeyouth}} (FB-app, hereafter) after approximately one year.
The initial survey consisted of advanced demographic questions regarding social issues and two validated psychometric questionnaires, one for personality assessment \cite{Costa1992} and one for morality assessment \cite{Haidt2007,Haidt2004}.
By visiting the FB-app the participants could fill in a number of questionnaires including the aforementioned validated questionnaires for personality and morality used in the initial survey. 
%The attributes, both demographic and psychometric, exploited in this study are chosen based on their utility for population segmentation, personalisation in communication for a wide range of topics. 
The above design allowed us to assess not only the demographic but also the psychometric biases,  due to (i) \textit{population}, comparing the attributes of people that engage or not on the specific platform, (ii) \textit{self-selection} during the recruitment phase where some participants chose not to respond, as well as (iii) \textit{behavioural}, due to the self-reporting variability probably due to the platform in which the questionnaire is administered.

In this study, we take a step forward towards raising awareness of the aforementioned biases entailed to the usage of a specific social platform for social science research with respect to a wide range of demographic and psychometric attributes.
Our findings suggest that surveys administered on Facebook do not exhibit major biases with respect to traditionally administered surveys, neither in terms of demographics nor personality attribute assessment.
From our analysis some small but statistically significant differences emerged which call for the researchers' attention.
In particular, Facebook platform introduces some population biases; participants without a Facebook profile appear to be less extroverted and more conscientious.
Regarding the self-selection bias at the recruitment phase, there were no significant differences neither for demographic nor for psychometric attributes.
The within-individual variability in responding was found to be low, indicating consistency in self-responding across all attributes between the Traditional and Facebook-administered survey;  nonetheless, a few behavioural biases emerged.
Comparing the overall population's behaviour in the two surveys loyalty, authority, and social binding values were higher, signalling that engaging in a social media platform slightly affects the individuals' behaviour.
Provided that the above biases are taken into consideration when designing research studies, our findings support the premise that Facebook is an adequate tool to administer surveys.

\section{Related Literature}

A review of the related literature shows that few studies have assessed the \textit{population}, \textit{self-selection} and \textit{behavioural} biases \cite{Schober2016}.
In a pioneering study on the topic, Zhao et al. \cite{Zhao2011} empirically compare the content of Twitter with a traditional news medium, New York Times, using unsupervised topic modelling.
Researchers mainly focus on Twitter data, addressing whether offline data can provide with the same findings as social media data \cite{Tufekci2014,Diaz2016}.
Diaz et al. \cite{Diaz2016} assuming that online and social media data is the output of some hypothetical pseudo-survey methodology, present an extended study, employing Twitter data, on how would this methodology would differ from conventional survey techniques.  
Tufekci \cite{Tufekci2014} raise a number of issues regarding the representativeness and validity of conclusions drawn based on social media analyses, stressing among other issues the preponderance of a single platform, Twitter,  mostly due to data and tools availability.

Moving from social media data to Internet obtained data in general, several scientific studies assessed the validity of results obtained from traditional versus online surveys administered on crowdsourcing platforms, for instance, Amazon's Mechanical Turk (AMT), \cite{Crump2013,Mason2012,Law2017,Baker2016}.
Their focus was on psychological and cognitive experiments \cite{Germine2012}, and all reached the conclusion that online crowdsourcing platforms consist of useful, feasible and desirable tools for research.
In particular, in one of the first systematic attempts to evaluate online questionnaires, Behrend et al.~\cite{Behrend2011} compare the behaviour of a sample of 270 participants from a crowdsourced environment as Amazon Mechanical Turk (AMT) with a similar sample from a traditional questionnaire answered by university students. They show that the crowdsourced sample had more demographic diversity than the university sample and their answers had better internal consistency.
Germine et al. \cite{Germine2012} address the data quality across a range of cognitive and perceptual tests applied to 25,000 participants in a Web environment. For three key performance metrics - mean performance, performance variance, and internal reliability - they observe that the results from self-selected Web samples do not differ systematically from those obtained from traditionally recruited and/or lab-tested samples.
Crump et al.~\cite{Crump2013}  conduct thoroughly designed experiments on AMT and observe that even for extended experiments that required problem-solving, learning, and precise millisecond control for response collection and stimulus presentation, the results seemed mostly in line with those of laboratory settings, as long as the experimental methods were solid.
Mason and Suri~\cite{Mason2012}  reach the same conclusion providing an analysis of the validity of the results obtained by crowdsourcing on the AMT platform. These studies together with Buhrmester et al. \cite{Buhrmester2011}, who points out that ``the data obtained are at least as reliable as those obtained via traditional methods'', support the use AMT as a platform for cognitive behavioural research.
Gosling and Manson~\cite{gosling2015internet} stress the Internet's future impact on the psychological research discussing aspects such as sample bias, anonymity, and ethical issues.
Baker et al. \cite{Baker2016}  assess the demographic characteristics of a large sample from a crowdsourced study to demonstrate how crowdsourcing can be used as an effective forensic research tool; their analyses, coupled with previous research on characteristics of crowdsourced samples, clearly indicate that crowdsourced samples are likely to be as adequate a source of clean interpretable data as university samples.
Finally, Law et al. \cite{Law2017} present a thorough analysis of the numerous crowdsourcing platforms, examining the circumstances under which crowdsourcing is a useful, feasible and desirable tool for research, as well as the factors that may influence researchers' decisions around adopting crowdsourcing technology.
Interestingly, all the above studies conclude that online experiments provide results of similar quality to those obtained by means of traditional experiments.

Recently, Schober et al. \cite{Schober2016} present an extensive review discussing whether social media content can
be compared with measurements from sample surveys, and whether survey research can be supplemented by
less costly data mining of already-existing or ``found'' data.
They raise a series of questions on the trustworthiness of such approaches, underlining the need for deeper understanding of the principles for aligning findings from social media analyses and surveys, which form the bases for important policy
decisions.
This study aims to shed light exactly on this topic having as core contribution the empirical assessment of  \textit{population}, \textit{self-selection} and \textit{behavioural} biases, when comparing results obtained from traditional versus Facebook administered surveys including both demographic and psychometric attribute.

\begin{figure*}[!ht]
\centering
\caption{Assessing \emph{population bias of the Facebook platform for the Big5 attributes}. Mann-Whitney U test on the personality traits of those who stated to maintain an Facebook profile (denoted as ``On FB'') against those who did not (denoted as ``Off FB''). The null hypothesis represents that both distributions are similar. On Facebook, participants are shown to be more extroverted and less neurotic. The stars above indicate the statistical significance for the null hypothesis rejection. All effect sizes are negligible with $d<0.1$(see section~\ref{sec:methods}). Note that, ``$\star\star\star$'' for p-value $<$ 0.001,``$\star\star$'' for p-value $<$ 0.01, ``$\star$'' for p-value $<$ 0.05, and 
 ``$-$'' for no statistical significance observed.}

\subfloat[Agreeableness]{\includegraphics[scale=0.28]{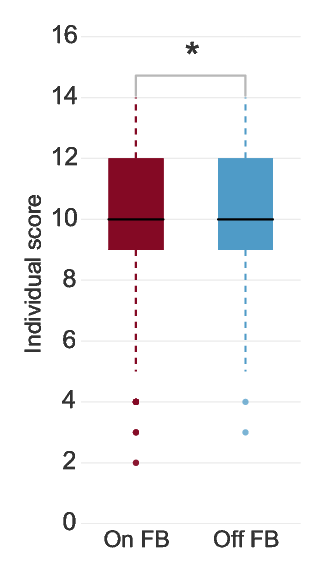}\label{fig:agree}}
\subfloat[Openness]{\includegraphics[scale=0.28]{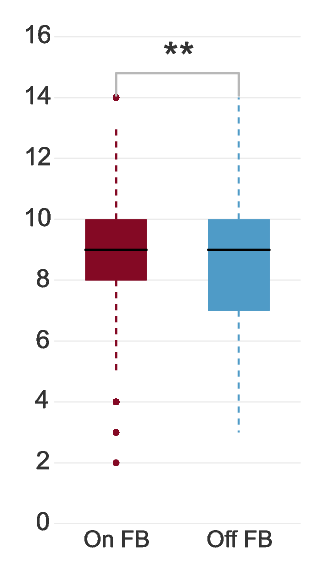}\label{fig:open}}
\subfloat[Extraversion]{\includegraphics[scale=0.28]{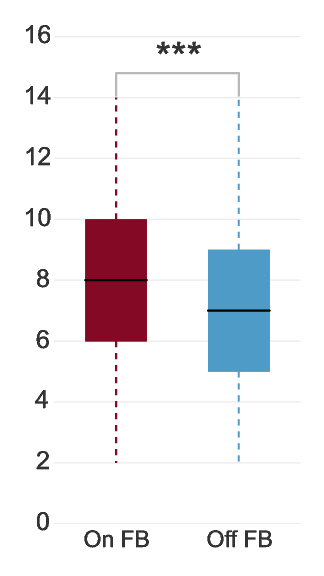}\label{fig:extra}}\\
\subfloat[Cons/ness]{\includegraphics[scale=0.28]{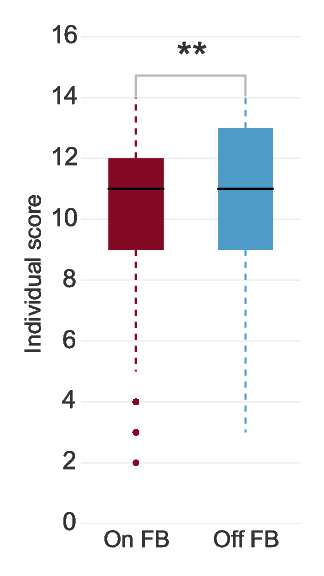}\label{fig:cons}}
\subfloat[Neurotisism]{\includegraphics[scale=0.28]{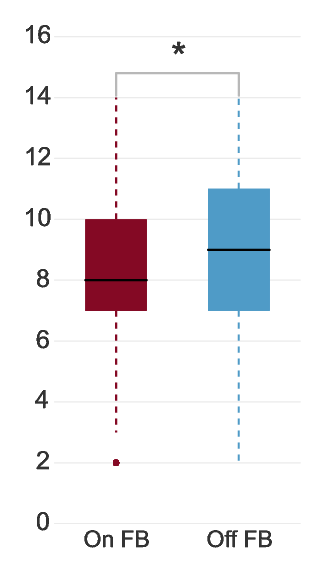}\label{fig:neuro}}
\label{fig:fb_big5}
\end{figure*}

\begin{figure*}[!ht]
\centering
\caption{Assessing \emph{population bias of the Facebook platform for the MFT attributes}. Mann-Whitney U test on the five moral foundations (\ref{fig:fb_care},\ref{fig:fb_fair},\ref{fig:fb_auth},\ref{fig:fb_loyal},\ref{fig:fb_pure}), including the two superior foundations, \textit{individualist} (\ref{fig:fb_individualist}), and \textit{binding}(\ref{fig:fb_binder}),
of those participants who stated to maintain a Facebook profile (denoted as ``On FB'') against those who did not (denoted as ``Off FB''). The null hypothesis represents that both distributions are the same. No statistically significant differences were found, i.e. we cannot reject the null hypothesis. All effect sizes are negligible with $d<0.1$(see section~\ref{sec:methods}).Note that, ``$\star\star\star$'' for p-value $<$ 0.001,``$\star\star$'' for p-value $<$ 0.01, ``$\star$'' for p-value $<$ 0.05, and 
 ``$-$'' for no statistical significance observed.}
\subfloat[Care]{\includegraphics[scale=0.28]{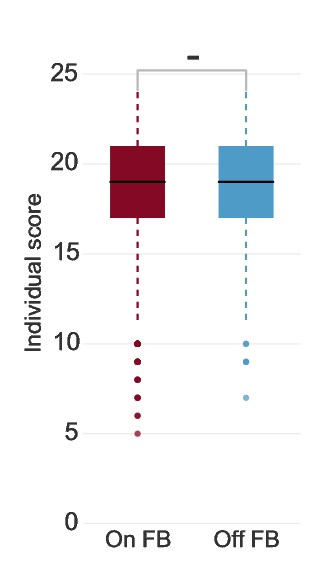}\label{fig:fb_care}}
\subfloat[Fairness]{\includegraphics[scale=0.28]{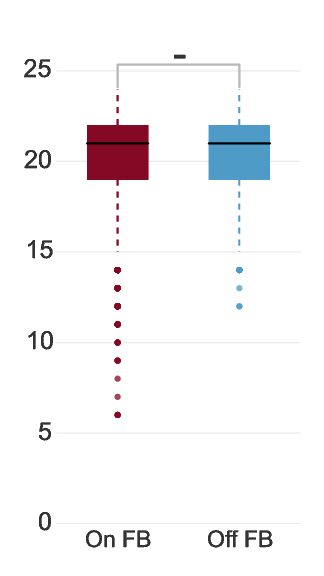}\label{fig:fb_fair}}
\subfloat[Authority]{\includegraphics[scale=0.28]{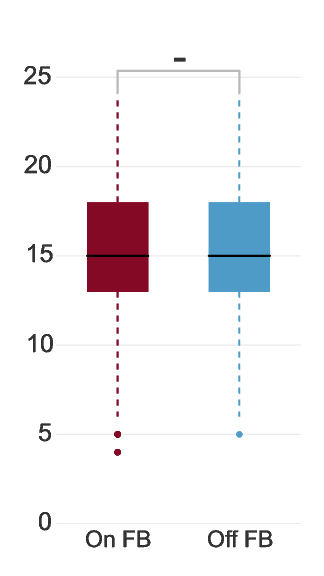}\label{fig:fb_auth}}
\subfloat[Purity]{\includegraphics[scale=0.28]{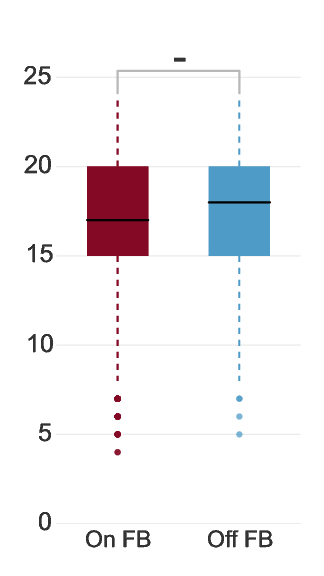}\label{fig:fb_loyal}}\\
\subfloat[Loyalty]{\includegraphics[scale=0.28]{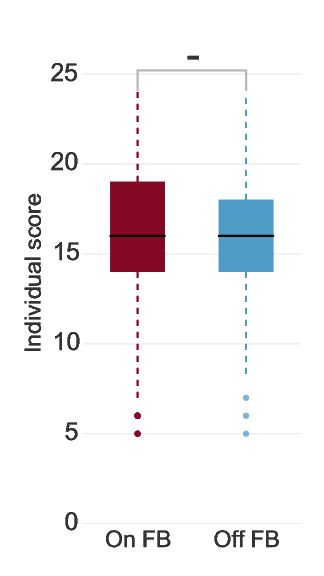}\label{fig:fb_pure}}
\subfloat[Individualism]{\includegraphics[scale=0.28]{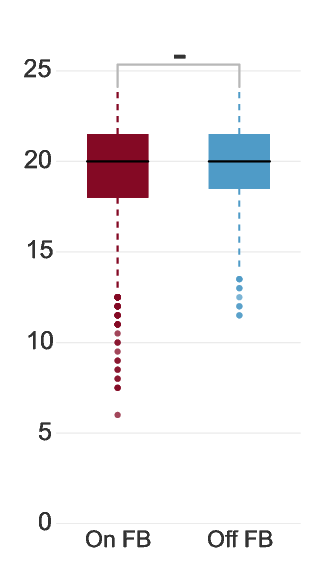}\label{fig:fb_individualist}}
\subfloat[Binding]{\includegraphics[scale=0.28]{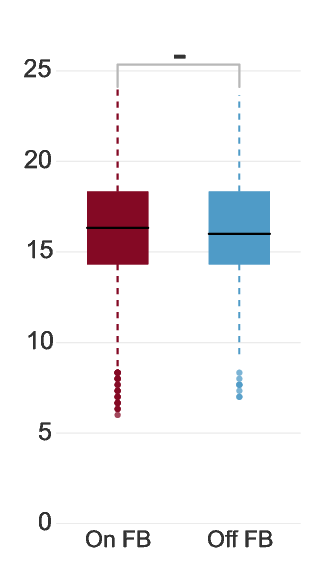}\label{fig:fb_binder}}
\label{fig:fb_mft}
\vspace{-10pt}
\end{figure*}

\section{Methods}
\label{sec:methods}

To address the scopes of this study, we designed an extended survey including a series of demographic variables, such as gender, age, geographical origin, education, employment and marital status\footnote{A link to the original survey will be provided after the blind review}.
Along with the demographic attributes we included two well-founded theories related to personality, moral values and combined are considered fundamental to the understanding of human decision-making processes and the individual's worldviews.
We included the Five Factor Inventory (Big5, hereafter) \cite{gosling2003very,Costa1992} for personality assessment and the Moral Foundations Questionnaire (MFQ, hereafter) \cite{Haidt2004,Haidt2007,graham2009liberals} for morality assessment.

%The assessment of personality and morality traits is of major importance since these attributes are often employed in crafting messages for personalised communication.
The Big5 personality traits model characterises personality based on five dimensions and has universal validity~\cite{schmitt2007geographic}.
In the following we describe its dimensions in terms of their two extremes:\\
\begin{itemize}
\item \noindent \textbf{Openness to experience}: inventive/curious vs. consistent/cautious.
\item \noindent \textbf{Conscientiousness}: efficient/organised vs. easy-going/careless.
\item \noindent \textbf{Extraversion}: outgoing/energetic vs. solitary/reserved.
\item \noindent \textbf{Agreeableness}: friendly/compassionate vs. analytical/detached.
\item \noindent \textbf{Neuroticism}: sensitive/nervous vs. secure/confident.
\end{itemize}

The Moral Foundations Theory (MFT) on the other hand focuses on the explanation of morality, its origins, development, and cultural variations \cite{Haidt2004,Haidt2007} and may be considered at a higher level, with respect to the dispositional traits of personality expressed in the Five Factor model \cite{Costa1992}. It provides insights on the characteristic adaptations of
the individuals \cite{Haidt2009} as described by Dan McAdams's three-level account of personality: (i) dispositional traits, (ii) characteristic adaptations, and (iii) life stories \cite{McAdams1995,McAdams2006}.
MFT focuses on the psychological basis of morality, identifying the following five moral foundations \cite{Haidt2007,Haidt2004}:\\
\begin{itemize}
\item \noindent \textbf{Care/Harm}: basic concerns for the suffering of others, including virtues of caring and compassion.
\item \noindent \textbf{Fairness/Cheating}: concerns about unfair treatment, inequality, and more abstract notions of justice.
\item \noindent \textbf{Loyalty/Betrayal}: concerns related to obligations of group membership, such as loyalty, self-sacrifice and vigilance against betrayal.
\item \noindent \textbf{Authority/Subversion}: concerns related to social order and the obligations of hierarchical relationships
such as obedience, respect, and proper role fulfilment.
\item \noindent \textbf{Purity/Degradation}: concerns about physical and spiritual contagion, including virtues of chastity, wholesomeness and control of desires.
\end{itemize}
The five moral foundations collapse into two superior ones, namely the (1) \textbf{individualising} and (2) \textbf{binding} foundations \cite{Haidt2007}. 
According to Haidt \cite{Haidt2007}, the individualising foundation asserts that the basic constructs of society are the individuals and hence focuses on their protection and fair treatment,
defending their right to pursue their own goals.  In contrast, the binding foundation focuses on group-binding, based on the respect of leadership and traditions, and the defence of the family as the nucleus of society.
The assessment of personality and morality traits is of major importance since these attributes are often fundamental explanatory factors tied to many sociological phenomena directly tied to demographic issues \cite{Bi2013,Kalimeri2017}. 
% add a few more here

To assess the differences between the populations in question since our data include both categorical and ordinal attributes we employed the Mann-Whitney U nonparametric statistical test.
The effect size is estimated as $d= \frac{2U}{mn} - 1$, where the two distributions are of size $n$ and $m$, where n and m are the population sizes and $U$ is the Mann-Whitney U statistic \cite{Cliff1993}.
We consider any effect size with magnitude $d$, as ``negligible''  if  $|d|<0.147$, ``small'' if $|d|<0.33$, ``medium'' if $|d|<0.474$ and ``large'' otherwise, according to the interpretation intervals suggested by \cite{Romano2006}.
The statistical significance level below which we can reject the null hypothesis and state that the distributions are different is depicted in the Figures as follows:
``$\star\star\star$'' for p-value $<$ 0.001,``$\star\star$'' for p-value $<$ 0.01, ``$\star$'' for p-value $<$ 0.05, and 
 ``$-$'' for no statistical significance observed.

\section{Experimental Design and Data Collection}

This study was conducted as part of a national-wide project launched in 2015 focusing on youth-related issues in Italy \cite{Blind}.
Our experiment was conducted in two phases with an approximately one-year time interval.
We employed a cohort which originates from the database of the aforementioned research project \cite{Blind} and is probability based and a demographically representative sample of the Italian youth population. 
The initial recruitment of this cohort was carried out by a mixed methodology, computer-assisted telephone interview (CATI), computer-assisted personal interview (CAPI) with in-depth computer-assisted web interview (CAWI), resulting in a sample of 9,358 individuals aged between 18 and 33 years (Mean = 25.7, STD = 4.7). 
%taking under consideration the age bracket as constituting emerging adulthood. The individuals were chosen with a stratified sampling technique. 
The cohort has been tested for representativeness with respect to a significant set of different variables, including gender, age, geographical origin, education, marital status, etc. (see \cite{Blind} for more details).
During the first stage, we invited via email the entire cohort to fill in a survey administered in a traditional web-based manner. 
This survey consisted of an extended number of questions regarding demographic and social issues related to youth and the aforementioned psychometric questionnaires about personality (Big5) and morality (MFQ).

Within a time interval of approximately one year from the initial survey, the cohort received an invitation via email to access our ad hoc Facebook-hosted application which, among other functions, administered the Big5 and MFQ psychometric questionnaires.
A consent form was obtained in terms of a privacy agreement which the participants declared to accept upon registration.
This procedure allowed us to account for three possible categories of biases as seen in the Table~\ref{tab:exp}.

%We opted for the Facebook platform based upon a few key criteria; first and foremost, its popularity \cite{Pew2018} permits the communication with a wide audience previously inaccessible; and secondly, the possibility of aligning surveys with longitudinal digital behavioural data, which is of extreme interest to the computational social science field.

%The attributes, both demographic and psychometric, exploited in this study are chosen based on their utility for population segmentation, personalisation in communication for a wide range of topics. 
%A demographically representative cohort of approximately 4,000 young - 17 to 33 years of age - adults in Italy, was engaged to participate in both phases of the experiment, filling in the survey.

\begin{table}[!h]
\renewcommand{\arraystretch}{1.2}
\centering
\begin{tabularx}{\columnwidth}{l l }
\toprule
Study~1. & \emph{Population Bias - Platform}\\
\midrule
& Comparison of the population differences of participants\\
&  who declared to maintain a Facebook account (On-FB) vs \\
& participants without a Facebook account (Off-FB).\\~\\
\hline
Study~2. & \emph{Self-Selection Bias - Recruitment}\\
\midrule
& Comparison of the population differences between the \\
&\textit{Traditional} cohort, those who refused to enter the FB-app \\
&(Invitees) versus the \textit{FB} cohort, those who accepted to enter\\
&   the FB-app (Recruited).\\~\\
\hline
Study~3. & \emph{Behavioural Bias} \\
\midrule
& Comparison of the within individual differences in responses  \\
& given in the traditional survey vs the FB-app\\
\bottomrule
\end{tabularx}
\caption{Overview of the analysis and scopes of each study. We focus on three types of possible bias namely, (i) population bias, due to the platform (Study~1. On/Off Facebook), (ii) to the self-selection (Study~2. \textit{Traditional} vs Facebook-administered survey), and (iii) the behavioural bias, due to individual differences between the \textit{Traditional} and the Facebook administered survey (Study~3.).}
\label{tab:exp}
\end{table}

The traditionally administered survey was filled in by 6,380 participants. To assess the quality of the data, we applied two simple criteria; participants with (i) identical responses to both Big5 and MFT individual questionnaire items, or (ii) mistaken responses in the two quality control questions, were excluded from the study.
After this preprocessing step we excluded approximately 34\% of the initial population; of an initial sample of 6,380 individuals, we remained with 4,239 participants which consist our \textit{Initial} cohort (see Table~\ref{tab:stats}).
After our email invitation approximately one year after the initial survey, approximately  76\% of the \textit{Initial} cohort did not login to the FB-app (participated only in the initial online survey) and are denoted as \textit{Traditional} cohort hereafter. The remaining 23\% of the \textit{Initial} cohort instead logged in the application, and are denoted as \textit{FB} cohort hereafter.

\section{Results and Discussion}

Table~\ref{tab:stats} reports the statistics on the two populations along with their demographic characteristics.
For each demographic attribute, we compared the \textit{Traditional}  and  \textit{FB} cohorts against the \textit{Initial} cohort.
No significant differences emerged, for any of the attributes, providing evidence of the fact that both the  \textit{Traditional} and the \textit{FB} cohorts are demographically representative subsets of the \textit{Initial} cohort with respect to age, gender, employment and educational level. The educational level is considered to be ``High'' if the participant declared at least to pursue a university degree and ``Low'' otherwise.

\begin{table}[!h]

\centering
\renewcommand{\arraystretch}{1.2}

\begin{tabular*}{\columnwidth}{l@{\extracolsep{\fill}} c@{\extracolsep{\fill}} c@{\extracolsep{\fill}} c@{\extracolsep{\fill}}}
\toprule
& Initial & Traditional & FB \\
\midrule
Population & 4,239 &3,251 & 956 \\
Age (std)  & 27.0 (4.2)& 27.1 (4.3)&26.9 (4.2) \\
Gender (Males)     & 65.3\% & 64.3\% &  65.8\%  \\
Employed (Yes) & 50.8\% & 50.4\% &50.9\% \\
Education (High) & 54.1\% & 55.2\% &54.4\% \\
\bottomrule
\end{tabular*}
\caption{Population size for the \textit{Initial} cohort as well as its two subsets, participants who filled in the online survey but chose not to enter the FB-app (\textit{Traditional}), and those who filled in the online survey and then participated also in the Facebook-administered one (\textit{FB}). For the three cohorts we present a comparison of the basic demographic information in terms of population, average and standard deviation of age, gender distribution of the population in terms of the percentage of male participants, the percentage of the population who is employed, and the level of education referring to the percentage of participants who have a university degree of any level or are enrolled in the university. Mann-Whitney U tests on all variables did not detect any statistically significant differences.}
\label{tab:stats}
\end{table}

\subsubsection{Study~1. Population Bias - Platform.}
As ``On-FB'',  we denote the participants who declared to maintain a Facebook profile while as ``Off-FB'' those who do not. Table~\ref{tab:fb_stats} reports the total number of participants in both populations as well as their demographic information.
We compared the two populations, ``On-FB''  and ``Off-FB'', according to the self-reported information they provided in the initial survey as for their demographic and psychometric attributes by means of Mann-Whitney U test.

The outcome of the test showed a statistical significant difference in the age of the two cohorts, with the \textit{FB} one to represent a slightly younger population (p-value $ < 0.001$ and $|d|<0.17$), while no other difference in demographic attributes was pointed out as statistically significant (see Table \ref{tab:fb_stats}).
Regarding their personality and moral traits we depict the obtained results in Figures \ref{fig:fb_big5} and \ref{fig:fb_mft}, respectively.
In Figure \ref{fig:fb_big5} and \ref{fig:fb_mft}, the coloured boxes represent the interquartile ranges while the median is depicted as a thick black horizontal line.
The dashed coloured lines represent the upper and lower quartile (whiskers) while the dots are the outliers of the distribution.
We observe that participants without a Facebook profile resulted as less extroverted (p-value $< 0.001$, $|d|<0.14$) with minor differences also present in other traits, as for example, their lower level of openness to new experiences (p-value $ < 0.001$, $|d|<0.1$), while at the same time they appear to be more conscientious p-value $ < 0.001$, $|d|<0.1$ and more neurotic (p-value $ < 0.01$, $|d|<0.1$).
Despite the limited size of the ``Off-FB'' sample - only 8.4\% of the cohort claimed not to have a Facebook profile - these differences are statistically significant and hence, should be considered. No significant differences were found for the moral domain attributes instead.

\begin{table}[!h]
\centering
\begin{tabular*}{\columnwidth}{l@{\extracolsep{\fill}} c@{\extracolsep{\fill}} c@{\extracolsep{\fill}}}
\toprule
& On FB  &  Off FB \\
\midrule
$\#$ of Participants & 3,882 & 357 \\
Age (std) & 26.9 (4.2)& 28.1 (4.0) \\
Gender (Males) & 65.8\% & 59.9\%\\
Employed (Yes) & 50.9\% &49.0\% \\
Education (High) & 54.4\% & 51.8\%\\
\bottomrule
\end{tabular*}
\caption{Descriptive statistics of the population that declared to maintain a Facebook profile (``On FB'') and the one that does not (``Off FB'').
For the three cohorts we present a comparison of the basic demographic information in terms of population, average and standard deviation of age, gender distribution of the population in terms of the percentage of male participants, the percentage of the population who is employed, and the level of education referring to the percentage of participants who pursue or have a university degree. 
A Mann-Whitney U test on the two cohorts pointed out only the age difference as statistically significant with $p < 0.001$ and small effect size $|d|<0.17$.}
\label{tab:fb_stats}
\end{table}

\subsubsection{Study~2. Self-selection Bias - Recruitment.}
\begin{figure*}[!ht]
\centering
\caption{\emph{Assessing \textit{self-selection} bias due to recruitment for Big5}. From the Mann-Whitney U test on the personality traits of those who participated in the survey but did not access the Facebook application (denoted as \textit{Traditional}) against those who accessed the application (denoted as \textit{FB}), no statistically significant differences are observed. Note that, ``$\star\star\star$'' for p-value $<$ 0.001,``$\star\star$'' for p-value $<$ 0.01, ``$\star$'' for p-value $<$ 0.05, and ``$-$'' for no statistical significance observed.}

\subfloat[Agreeableness]{\includegraphics[scale=0.28]{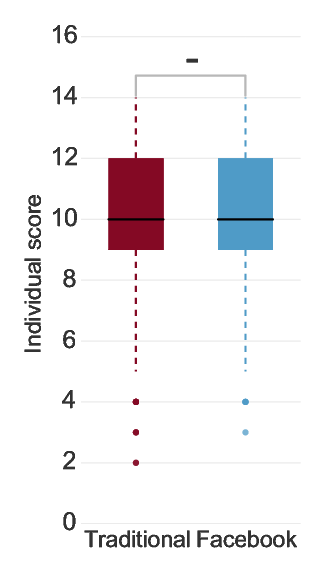}\label{fig:agree}}
\subfloat[Openness]{\includegraphics[scale=0.28]{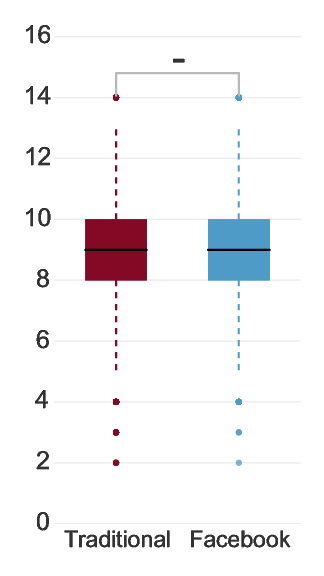}\label{fig:open}}
\subfloat[Extraversion]{\includegraphics[scale=0.28]{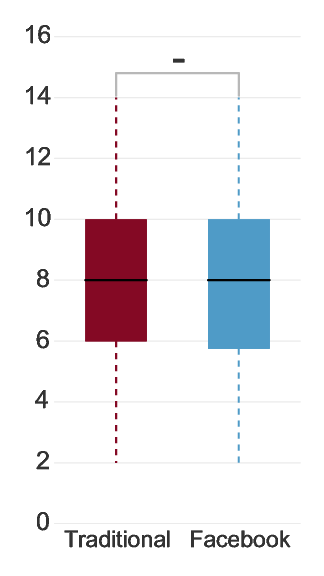}\label{fig:extra}}\\
\subfloat[Cons/ness]{\includegraphics[scale=0.28]{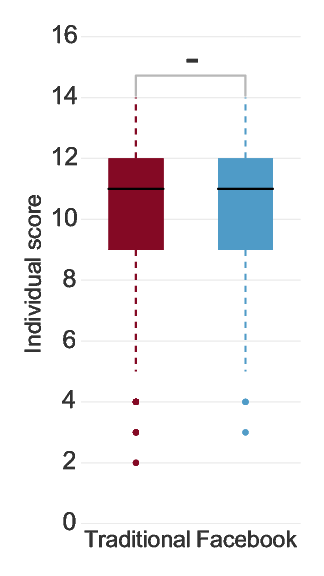}\label{fig:cons}}
\subfloat[Neuroticism]{\includegraphics[scale=0.28]{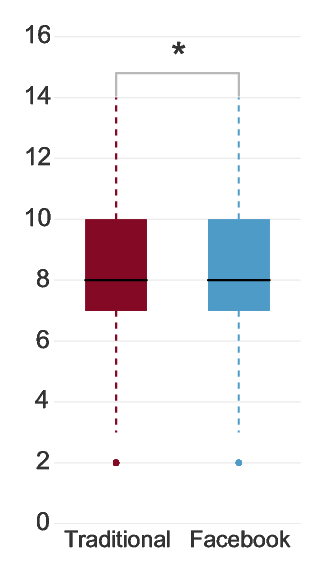}\label{fig:neuro}}
\label{fig:login_big5}
\vspace{-10pt}
\end{figure*}

\begin{figure*}[!ht]
\centering
\caption{\emph{Assessing \textit{self-selection} bias due to recruitment for the MFT}. Mann-Whitney U test on the five basic moral foundations (MFT) of those who participated in the survey but did not access the Facebook application (denoted as \textit{Traditional}) against those who accessed the application (denoted as \textit{FB}). The stars indicate the level of the p-value.
We note that in the \textit{FB} group tends to be lower in Purity, Loyalty, and Social Binding, however, the differences are small. For all the attributes with statistically significant effects, the effect sizes are negligible with $d<0.14$ (see section~\ref{sec:methods}). Note that, ``$\star\star\star$'' for p-value $<$ 0.001,``$\star\star$'' for p-value $<$ 0.01, ``$\star$'' for p-value $<$ 0.05, and 
 ``$-$'' for no statistical significance observed.}
\subfloat[Care]{\includegraphics[scale=0.28]{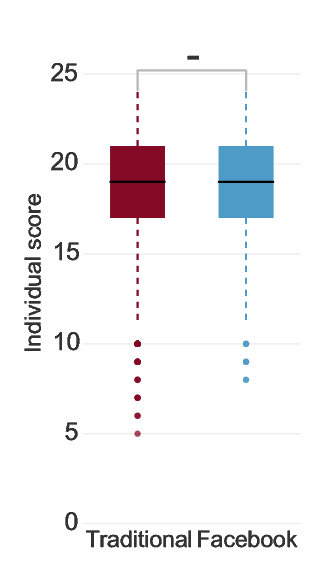}\label{fig:login_mft_care}}
\subfloat[Fairness]{\includegraphics[scale=0.28]{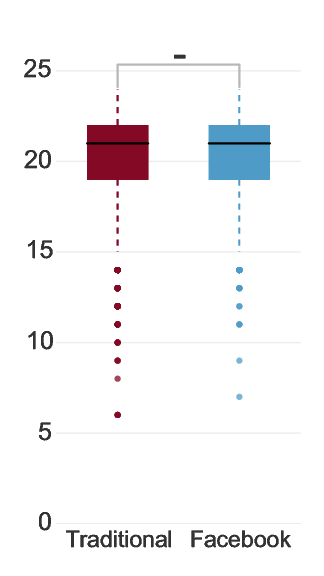}\label{fig:login_mft_fair}}
\subfloat[Authority]{\includegraphics[scale=0.28]{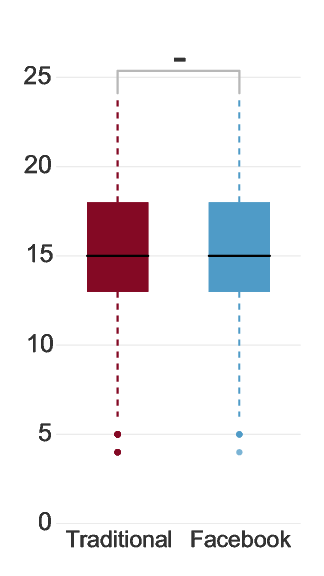}\label{fig:login_mft_auth}}
\subfloat[Purity]{\includegraphics[scale=0.28]{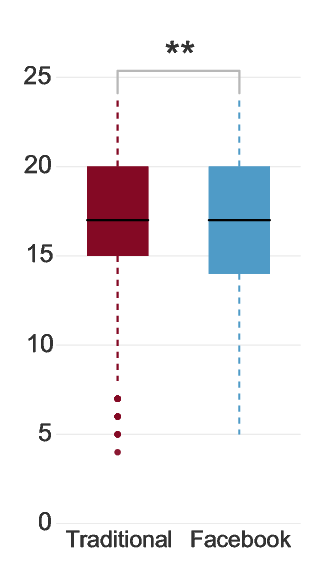}\label{fig:login_mft_loyal}}\\
\subfloat[Loyalty]{\includegraphics[scale=0.28]{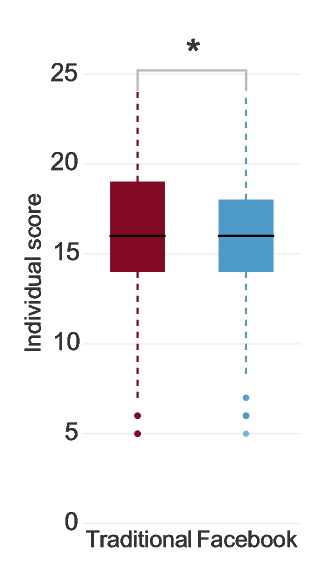}\label{fig:login_mft_pure}}
\subfloat[Individualism]{\includegraphics[scale=0.28]{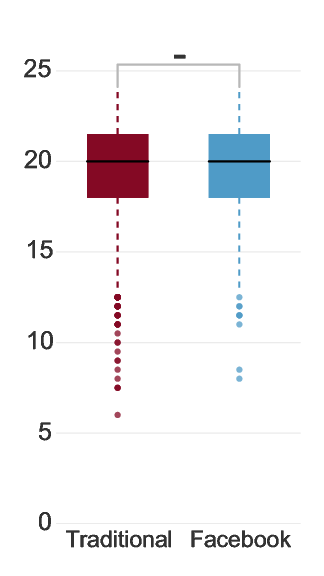}\label{fig:login_mft_individualist}}
\subfloat[Binding]{\includegraphics[scale=0.28]{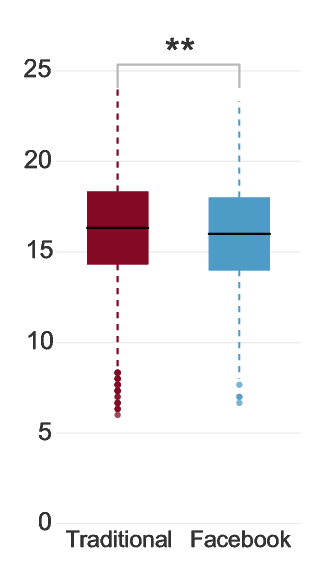}\label{fig:login_mft_binder}}
\label{fig:login_mft}
\end{figure*}

To assess the self-selection biases in the recruitment phase, we compared the demographic and psychometric attributes of the participants in the \textit{Traditional} cohort against those in the \textit{FB} cohort by means of Mann-Whitney U test.
To avoid any confounding factors introduced by the platform, we compared the responses obtained from the initial survey for both groups.
The Mann-Whitney U test did not show any significant differences neither for the demographic attributes (see Table \ref{tab:fb_stats}) nor for the Big5 personality traits (see Figure \ref{fig:login_big5}); It did however point out differences in the purity (p-value $ < 0.001$, $|d|<0.06$), loyalty (p-value $< 0.01$, $|d|<0.09$)  and binding (p-value $ < 0.001$, $|d|<0.06$) values, all with negligible effect size (see Figures~\ref{fig:login_mft_pure}, \ref{fig:login_mft_loyal}, \ref{fig:login_mft_binder}).

\subsubsection{Study~3. Behavioural Bias.}

\begin{figure*}[!ht]
\centering
\caption{\emph{Assessing \textit{behavioural} bias for the Big5}. Mann-Whitney U test on the personality traits of those who participated in the initial survey and then accessed the Facebook application. The comparison is made on the responses to the traditional  survey and ones given to the Facebook administered one. No statistically significant differences are noted between the distributions.}
\subfloat[Agreeableness]{\includegraphics[scale=0.28]{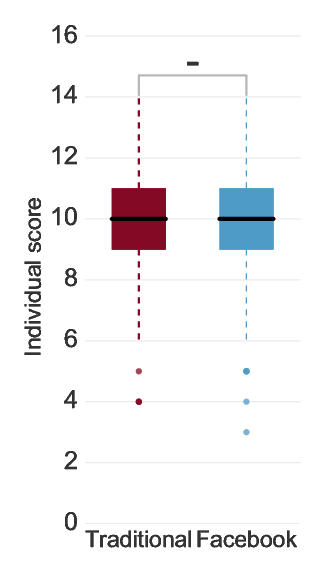}\label{fig:agree}}
\subfloat[Openness]{\includegraphics[scale=0.28]{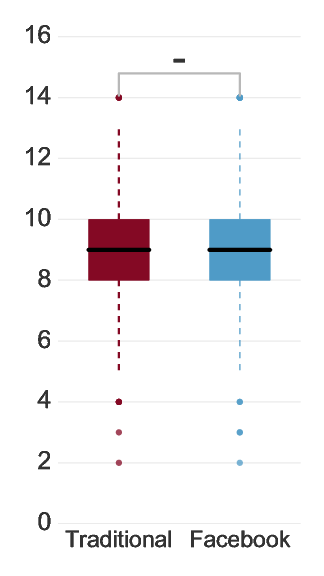}\label{fig:open}}
\subfloat[Extraversion]{\includegraphics[scale=0.28]{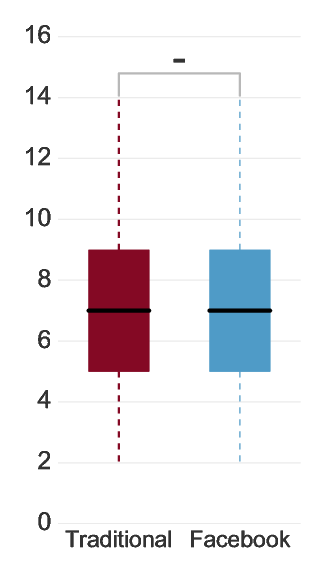}\label{fig:extra}}\\
\subfloat[Cons/ness]{\includegraphics[scale=0.28]{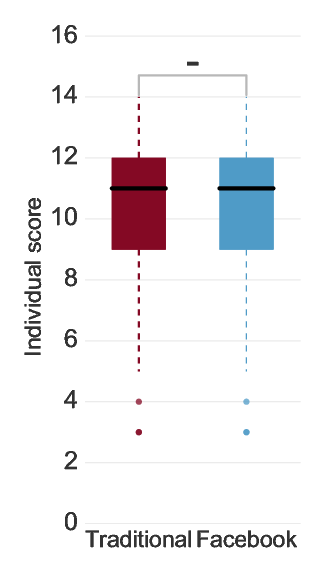}\label{fig:cons}}
\subfloat[Neuroticism]{\includegraphics[scale=0.28]{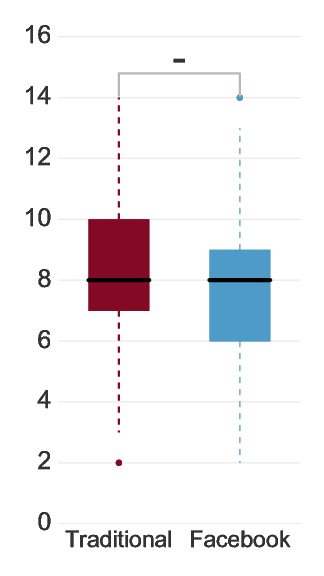}\label{fig:neuro}}
\vspace{-10pt}
\label{fig:ly_big5}
\end{figure*}

\begin{figure*}[!ht]
\centering
\caption{\emph{Assessing \textit{behavioural} bias for the MFT}. Mann-Whitney U test on the moral foundations of those who participated in the traditional survey and then accessed the Facebook application. The comparison is made on the responses to the traditional  survey and ones given to the Facebook administered one. Individuals rated themselves higher in authority, loyalty and social binding foundations within the Facebook administered survey, which is probably an effect of the platform's social character. Stars indicate the statistical significance. All effect sizes are negligible with $d<0.1$  (see section~\ref{sec:methods}).}

\subfloat[Care]{\includegraphics[scale=0.28]{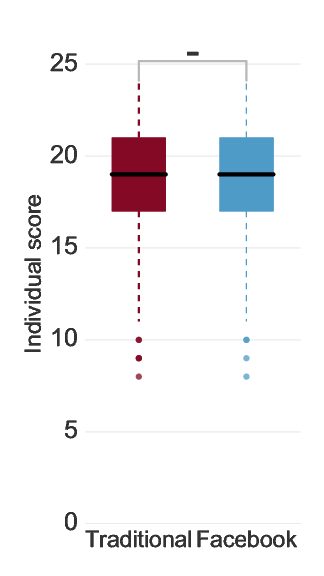}\label{fig:care}}
\subfloat[Fairness]{\includegraphics[scale=0.28]{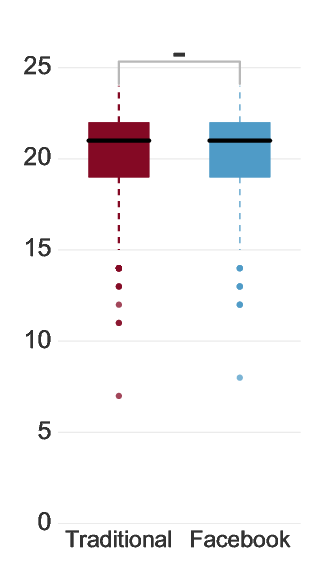}\label{fig:fair}}
\subfloat[Authority]{\includegraphics[scale=0.28]{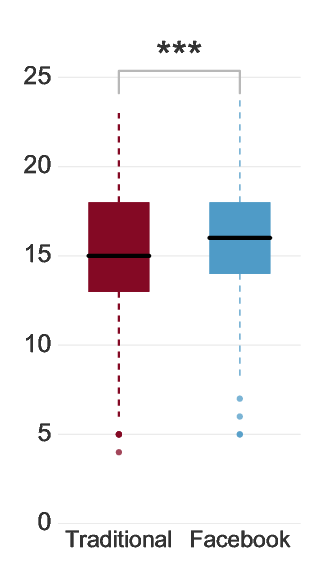}\label{fig:auth}}
\subfloat[Purity]{\includegraphics[scale=0.28]{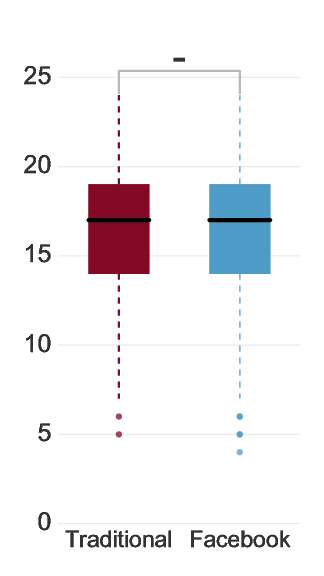}\label{fig:loyal}}\\
\subfloat[Loyalty]{\includegraphics[scale=0.28]{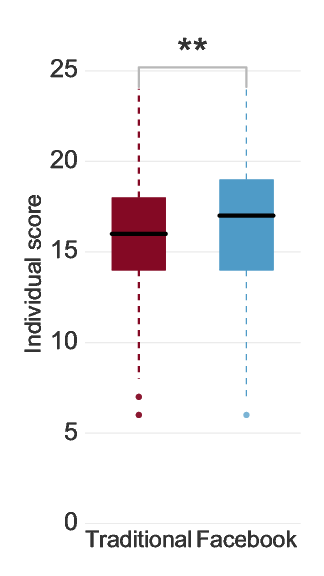}\label{fig:pure}}
\subfloat[Individualism]{\includegraphics[scale=0.28]{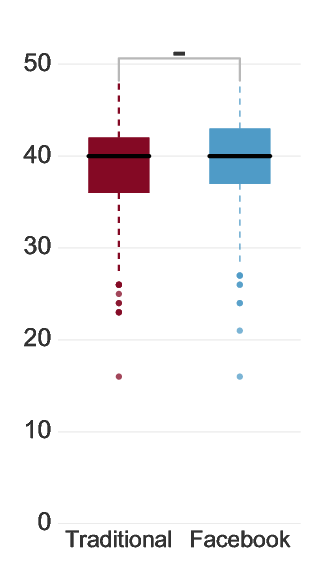}\label{fig:individualist}}
\subfloat[Binding]{\includegraphics[scale=0.28]{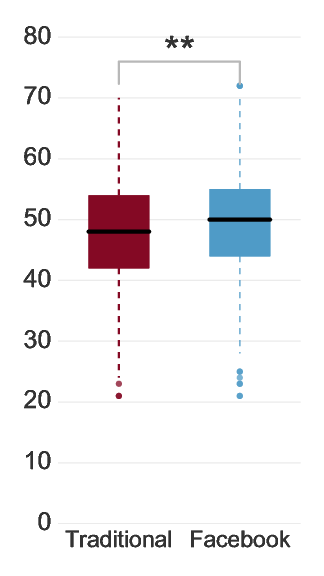}\label{fig:binder}}
\label{fig:ly_mft}
\end{figure*}

Finally, we focused on behavioural biases due to self-reporting; to do so, we compared the participants' responses in the traditional survey against their responses on the Facebook administered one.
The analysis showed that they when responding in the FB-app they judged themselves as slightly more authoritarian (p-value $< 0.001$, $|d|<0.14$) and loyal (p-value $ < 0.001$, $|d|<0.11$) (Figures \ref{fig:ly_big5} and \ref{fig:ly_mft}). 
They also claimed to value more social binding principles (p-value $ < 0.001$, $|d|<0.11$). These latter findings may due to Facebook's intrinsic social character.

\begin{figure*}[!ht]
\centering
\caption{Individual self-consistency of personality traits (Big5) reporting evaluated by means of Kendall's Tau correlation between an individual's responses on the traditional survey and on Facebook (indicated by the blue dot). The box plot shows the median and interquartile range of the Kendall's Tau correlation values obtained when we shuffle the responses in the  \textit{Traditional} and \textit{FB} cohorts for all the individuals 1,000 times. The difference between the actual correlation value and the randomised experiment, sustains the claim for self-consistency between traditional and  Facebook administered surveys.}
\subfloat[Agreeableness]{\includegraphics[scale=0.28]{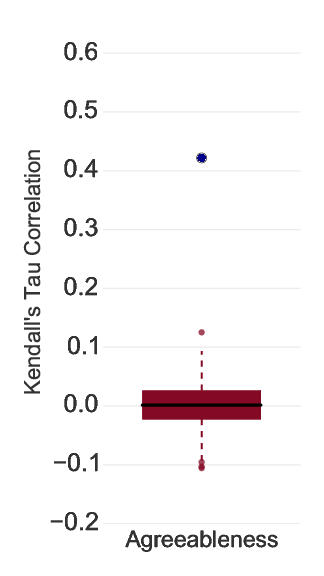}\label{fig:agree}}
\subfloat[Openness]{\includegraphics[scale=0.28]{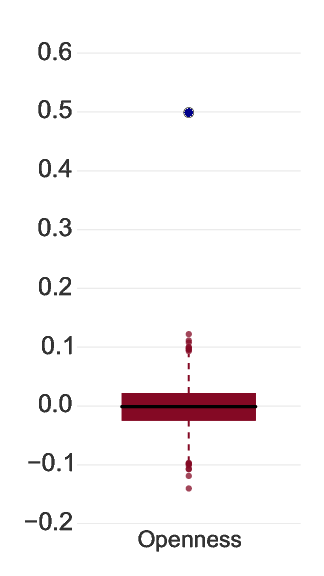}\label{fig:open}}
\subfloat[Extraversion]{\includegraphics[scale=0.28]{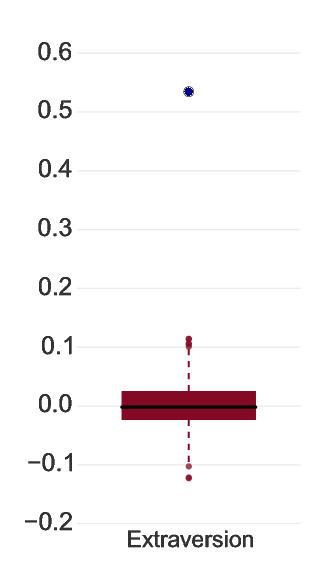}\label{fig:extra}}\\
\subfloat[Cons/ness]{\includegraphics[scale=0.28]{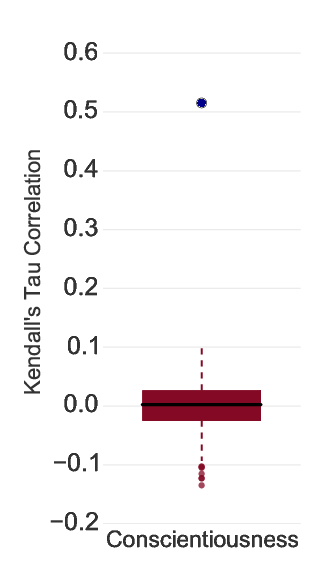}\label{fig:cons}}
\subfloat[Neuroticism]{\includegraphics[scale=0.28]{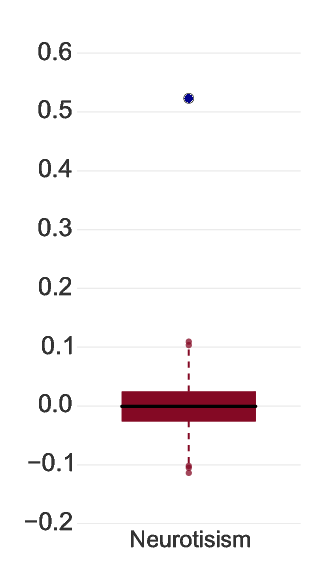}\label{fig:neuro}}
\label{fig:corr_big5}
\end{figure*}

\begin{figure*}[!ht]
\centering
\caption{Individual self-consistency of moral foundations (MFT)  reporting evaluated by means of Kendall's Tau correlation between an individual's responses on the traditional survey and on Facebook (indicated by the blue dot). The box plot shows the median and interquartile range of the Kendall's Tau correlation values obtained when we shuffle the responses in the  \textit{Traditional} and \textit{FB} cohorts for all the individuals 1,000 times. The difference between the actual correlation value and the randomised experiment, sustains the claim for self-consistency between traditional and  Facebook administered surveys.}
\subfloat[Care]{\includegraphics[scale=0.28]{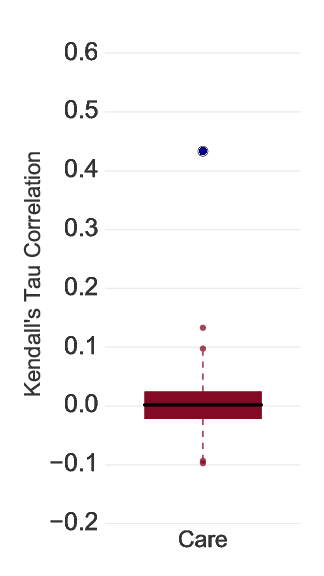}\label{fig:care}}
\subfloat[Fairness]{\includegraphics[scale=0.28]{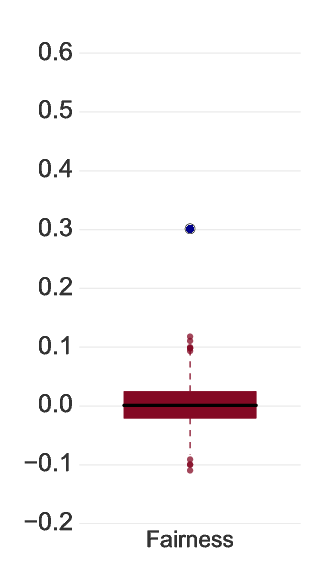}\label{fig:fair}}
\subfloat[Authority]{\includegraphics[scale=0.28]{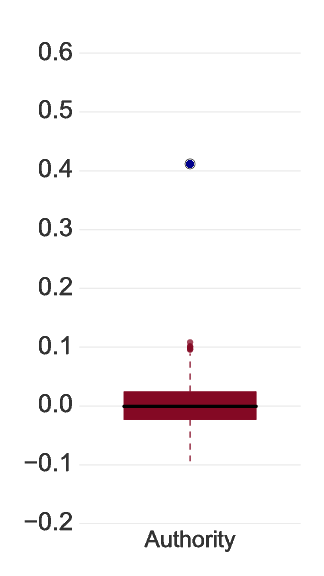}\label{fig:auth}}
\subfloat[Purity]{\includegraphics[scale=0.28]{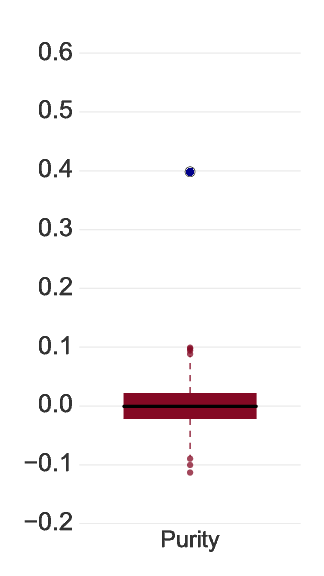}\label{fig:loyal}}\\
\subfloat[Loyalty]{\includegraphics[scale=0.28]{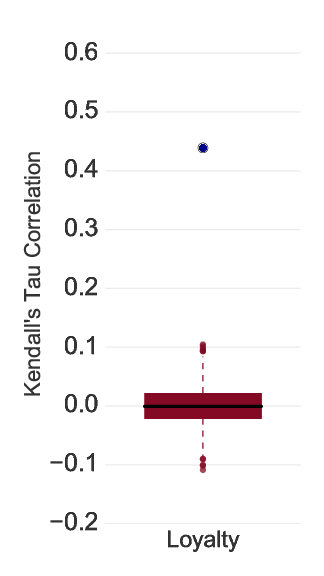}\label{fig:pure}}
\subfloat[Individualism]{\includegraphics[scale=0.28]{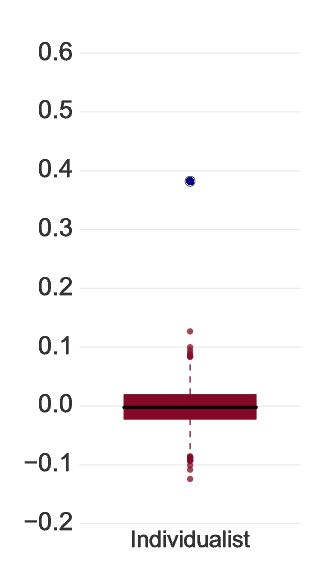}}
\subfloat[Binding]{\includegraphics[scale=0.28]{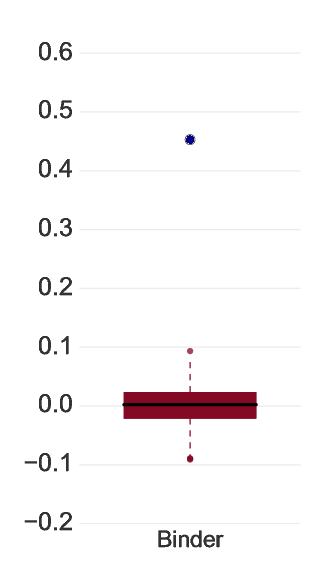}\label{fig:binder}}
\label{fig:corr_mft}
\end{figure*}

To assess the within-subject variability we compared the psychometric scores the participants reported in the traditional survey and the ones they reported within our FB-app.
For each attribute, we estimated the Kendall's Tau correlation values obtained from the individual responses in the traditional and the respective Facebook survey. The blue dots in Figures~\ref{fig:corr_big5} and ~\ref{fig:corr_mft} report the obtained results. 
Then, we randomly shuffled the answers of all participants  in the \textit{Traditional} and \textit{FB} cohorts 1,000 times and computed the Kendall's Tau correlation value each time. 
The median and interquartile ranges of the bootstrapped correlation distributions, between the individual responses on  the traditional and Facebook surveys are shown in the box plots of Figures~\ref{fig:corr_big5} and ~\ref{fig:corr_mft} for the personality and morality attributes, respectively. The correlations between the two surveys lie at an intermediate range (from 0.3 to 0.55), however,  significantly higher than the null model (see box plots in Figures~\ref{fig:corr_big5} and ~\ref{fig:corr_mft}). This supports the idea that there is good consistency in self-reporting.

\subsection{Limitations}
Understandably, this study entails a series of limitations; first and foremost, our sample is a youth population in Italy. Apart from the geographical and cultural effect, young people are of course more at ease with sharing their private information \cite{Anderson2010,Burkell2014,Kezer2016}. At the same time, since they have already participated in a traditional survey, they are accustomed to taking questionnaires regarding their personal and demographic attributes. 
Since our \textit{Initial} cohort is representative of the Italian youth and the recruited population on Facebook closely follows the same demographic characteristics, we claim that a participant that is recruited on Facebook follows the demographics of the population under investigation. 
We are only able to make claims about the people that are part of our cohort though, and we cannot draw conclusions on the average Facebook user; which does not fall under the scopes of this study. 
Given the limited size of our cohort (approximately 4,000 participants) our findings are to be interpreted with caution. 
Moreover, we acknowledge that initial survey might be subject to the same methodological biases of every survey \cite{Groves2010,Olteanu2016}, which, however, are beyond of our control; the same holds for the recruitment and the follow-up survey.

\section{Conclusions and Future Directions}

Since the 2000's, the massification of the Internet brought significant advantages to the collection of research data, in terms of enrichment and diversity of data, while at the same time reduced the research costs.
Social media can complement existing practices and provide new insights into demographic and social studies on population.
The core contribution of this study is the assessment of the biases entailed in these data, possibly due to subversive behaviours when participating in social media administered studies.
We focused on differences in demographic and psychometric attributes that might indicate (i) population, (ii) self-selection in the recruitment phase; and (iii) behavioural biases.

Considering our limited size of the cohort (4,000 people on the \textit{Initial} cohort and 988 people on the \textit{FB} one) and its focus on a specific geographic location and age range, our findings suggest that the population that chose not to proceed to the Facebook administered survey does not exhibit major biases with respect to the population of the traditionally administered one neither in terms of demographics, nor psychometric attributes.
Consequently, we conclude that our evidence supports the claim that self-selection biases of the Facebook platform are negligible. 

Conversely,  when carrying out surveys on Facebook \textit{population} and \textit{behavioural} biases are to be taken into account. 
In terms of \textit{population} biases, participants that declared not to maintain a Facebook profile resulted to be more introverted, conscientious and neurotic with respect to the ones that do use Facebook when analysing their personality traits (Big5).
Regarding \textit{behavioural} biases,  some small, yet statistically significant, behavioural differences emerged between the responses in the traditional and the Facebook administered surveys. When on Facebook participants rated themselves as more loyal, authoritarian and more fond of social binding values, which may indicate that engaging in a social media platform like Facebook slightly affects the individuals' behaviour reflected on their self-reporting.

This study contributes to the limited body of research on this arising issue, with an empirical assessment on \textit{population}, \textit{self-selection} and \textit{behavioural} biases present in surveys administered on social media.
The results obtained from traditional or Facebook administered surveys are of similar quality with respect to basic demographic and psychometric attributes. 
Moreover, given the cost-effectiveness of the platform, such surveying approaches can supplement the traditional demographic and sociological practices in addressing research questions timely and in greater scale. 
Keeping in mind the limitations of our study and the observed biases, our findings suggest that the Facebook platform can be employed as a valid research tool to administer social and psychometric research surveys, nonetheless, its not entirely neutral character should be considered to achieve impartiality.

\section{Acknowledgments}
For the figures we adapted the code developed by Dr. Jean-Baptiste Mouret, available here:
\href{https://github.com/jbmouret/matplotlib\_for\_papers}{Link}.
%We are grateful to Dr. Luca Rossi for the helpful discussions.


%\begin{small}
\bibliographystyle{plain}
%\bibliography{sample,DemographicResearch}
%\end{small}
\nolinenumbers

%This is where your bibliography is generated. Make sure that your .bib file is actually called library.bib

\end{document}